\documentclass[]{interact}

\usepackage{changepage}
\usepackage{epstopdf}
\usepackage{subfigure}
\usepackage{fixltx2e}

\theoremstyle{plain}

\theoremstyle{definition}

\theoremstyle{remark}

\begin{document}

\title{Unstable Footwear as a Speed-Dependent Noise-Based Training Gear to Exercise Inverted Pendulum Motion During Walking}

\author{
\name{F. Dierick \textsuperscript{a,b}\thanks{Email: dierickf@helha.be}, A.-F. Bouch\'e \textsuperscript{a}, M. Scohier \textsuperscript{a}, C. Guille \textsuperscript{a} and F. Buisseret \textsuperscript{a,c}}
\affil{\textsuperscript{a} Forme \& Fonctionnement Humain Research Unit, Haute Ecole Louvain en Hainaut, Rue Trieu Kaisin 134, Montignies sur Sambre, Belgium\\
 \textsuperscript{b}  Faculty of Motor Sciences, Universit\'e catholique de Louvain, Louvain-la-Neuve, Belgium\\
  \textsuperscript{c}  Service de Physique Nucl\'{e}aire et Subnucl\'{e}aire,
Universit\'{e} de Mons, UMONS  Research
Institute for Complex Systems, Place du Parc 20, 7000 Mons, Belgium}
}

\maketitle
\begin{abstract}
Previous research on unstable footwear has suggested that it may induce plantar mechanical noise during walking.
The purpose of this study was to explore whether unstable footwear could be considered as a noise-based training gear to exercise body center of mass (CoM) motion during walking or not.
Ground reaction forces were collected among 24 healthy young women walking at speeds between 3 and 6 km h\textsuperscript{-1} with control running shoes and unstable rocker-bottom shoes. The external mechanical work, the recovery of mechanical energy of the CoM during and within the step cycles, and the phase shift between potential and kinetic energy curves of the CoM were computed.
Our findings support the idea that unstable rocker-bottom footwear could serve as a speed-dependent noise-based training gear to exercise CoM motion during walking. At slow speed, it acts as a stochastic resonance or facilitator, whereas at brisk speed it acts as a constraint.\\
\end{abstract}


\begin{keywords}
Unstable footwear, Walking, Motion analysis, Stochastic noise
\end{keywords}


\section{Introduction}
Walking for leisure, health or fitness requires minimal equipment, footwear being the most important. While traditional footwear is designed to provide stability during the stance phase of walking, the concept of unstable footwear was born in 1996 with the introduction of the first Masai Barefoot Technology (MBT) shoe on the market. Since then, and given the high marketability of unstable footwear specifically in women, several other manufacturers introduced their own rocker-bottom or rounded pods models.

Unstable shoes are characterized on one hand by intentionally rounded outsoles in the anterior-posterior direction, which facilitate rockers of the foot, and on the other hand by a compressible rear foot section, which promotes``natural instability"~\cite{bib1}. Instability levels generated by rocking footwear are perfectly managed by the central nervous system but leads to an increased variability of walking, so much that this kind of footwear is considered as a motor constraint applicable during everyday activity and in particular during walking~\cite{bib2}. Nowadays, the addition of noise or variability during everyday activities is of great importance for coaches, therapists, and sport scientists since it can be used to enhance the quality of training, practice, and rehabilitation programs. Introduction of mechanical noise to the plantar mechanoreceptors may be a low-cost, simple and attractive training approach since the addition of noise to a nonlinear dynamics of a biological system induces stochastic resonance~\cite{bib3} that could have a positive impact on human postural control and gait~\cite{bib6,bib4,bib5}.

In human gait, conversion from kinetic energy of motion into gravitational potential energy during the first half of the stance phase and its returned as kinetic energy during the second half has been linked to that of an inverted pendulum~\cite{bib7}. The pivot point of the inverted pendulum is the ankle joint and foot and the motion of the ankle-foot complex relative to the ground is described as 3 or 4 functional rockers~\cite{bib8}: heel, ankle, forefoot, and toe. These rockers allow a heel-sole-toe rolling contact over the ground during each walking step that is very unusual outside the hominoidea~\cite{bib9}, with a forward translation of the center of pressure, roughly analogous to that of a wheel~\cite{bib10}.

Previous studies~\cite{bib11,bib7} showed that the recovery of mechanical energy of the center of mass (CoM) during the step cycle deviates from that of an ideal pendulum, with a maximal recovery value around 65\% at the optimal walking speed. Consequently, the remaining 35\% of mechanical energy necessary to achieve the pole-vault motion of the body's CoM must be actively performed by the musculoskeletal system and is usually quantified as the external mechanical work~\cite{bib12}. The existence of such a pendulum mechanism of gait is of major concern in a stochastic perspective since nonlinear structure of gait is dependent on the neural control of the forward progression of the CoM during the stance phase of gait~\cite{bib13}. Furthermore, an in-depth analysis of the instantaneous recovery of mechanical energy within each step has been previously developed and has been shown to be a powerful tool in understanding the pendulum mechanism of walking~\cite{bib14}, but this method has never be applied to the study of noise or variability induced by unstable footwear.

Since heel-sole-toe walking strategy has been implicated in the whole body mechanics~\cite{bib15} and that wearing unstable footwear with a rocker bottom imply a redistribution of the external mechanical work done during the step at the spontaneous walking speed~\cite{bib16}, we hypothesize that an unstable footwear with a rocker-bottom design could be used as a training gear allowing to modify the pendulum energy conversion mechanism during walking. To our best knowledge, and despite the recommendation to use unstable shoes as a motor constraint~\cite{bib2}, no study addressed this specific research topic. 

\section{Materials and Methods}
\subsection{Participants and footwear}

Twenty-four healthy young women (age = 20.4 $\pm$ 1.6 years, weight = 63.9 $\pm$ 8.6 kg, height = 167.3 $\pm$ 4.6 cm, european shoe size = 40 $\pm$ 1) were enrolled in the study and recruited among students of our physical therapy department. They were recreationally active, by which we mean a participation in physical activity for a minimum of 20 minutes, 3 times a week, but no engagement in competitive walking.

The participants had to meet the following inclusion criteria: being an adult female, aged of maximum 25 years, able to walk without limping and irregularities at walking speeds between 3 and 6 km h\textsuperscript{-1}, with no ankle joint instability. Moreover, it had been required that the participants did not have any injury or trauma to the lower limbs during the 3 months prior to the measurements. Before the experiments, the purpose and nature of the study were explained to the subjects and their informed written consent was obtained. The experiments were performed according to the Declaration of Helsinki and were approved by the ethics committee of Grand H\^opital de Charleroi. All subjects wore sportswear during experiments.

In this study we compared the walking of women with two different pairs of footwear: a pair of control footwear (CTRL) and a pair of unstable footwear (UNST). CTRL were classic running footwear (Ascot Trainers for Women) and UNST were footwear with rounded soles in the anterior-posterior direction (Pata Women's model, Massai Barefoot Technology, MBT). The characteristics of footwear are reported in Table~\ref{table1} and pictures in Fig. \ref{Fig. 1}A.

\subsection{Experimental procedure}
Walking measurements with both footwear were carried out on the same half day and lasted about one hour per participant. Before walking measurements, height and weight of each participant were collected.

First, participants wore CTRL footwear and were asked to walk naturally on a force platform (3.20 m long and 0.50 m wide) mounted at floor level and embedded in the center of a 13 m long walkway. The force platform was made of 4 separate strain-gauges plates of 0.80 m long (3D-Force Plate, Arsalis, Louvain-la-Neuve, Belgium) and ground reaction forces (GRFs) were collected in the three directions (vertical, forward, and lateral) at a sampling frequency of 1 kHz.

The participants had then to walk at 4 different target speeds of progression: 3, 4, 5 and 6 km h\textsuperscript{-1}. The average speed of progression was calculated using two pairs of photocells, located at both ends of the force platforms, and whose height was adjusted at the neck of the participant. Five walking trials were recorded for each speed of progression. A trial was considered valid when: (1) average speed of progression of the participant was at $\pm$ 0.3 km h\textsuperscript{-1} of the target speed, which corresponds to a maximal relative error of 10\% for walking speed in the considered range; and (2) average speed of progression of the participant was constant, with the criterion that vertical and forward accelerations of the body's CoM during one or more complete strides differed from the decelerations by less than 25\%~\cite{bib7}.

Second, participants wore UNST footwear. They were instructed by an investigator (CG) on ``proper" walking technique using a video and asked to follow a 20 min familiarization procedure that consisted of 10 minute walk on level ground and 10 minute walk on a motorized treadmill (Treadmill 2000, Marquette, Milwaukee, WI) with 1\% incline. After treadmill walking, participants were allowed to walk again freely on level ground during 10 min and the same investigator observed the walk of the participants to verify that they were well accustomed to walking with UNST footwear (observed variables were the fluidity of the ankle joint and foot kinematics, and the absence of limping). Finally, walking measurements were realized with UNST footwear in the same conditions as the CTRL footwear. 

\subsection{Data Processing}
\begin{figure}
\includegraphics[scale=0.6]{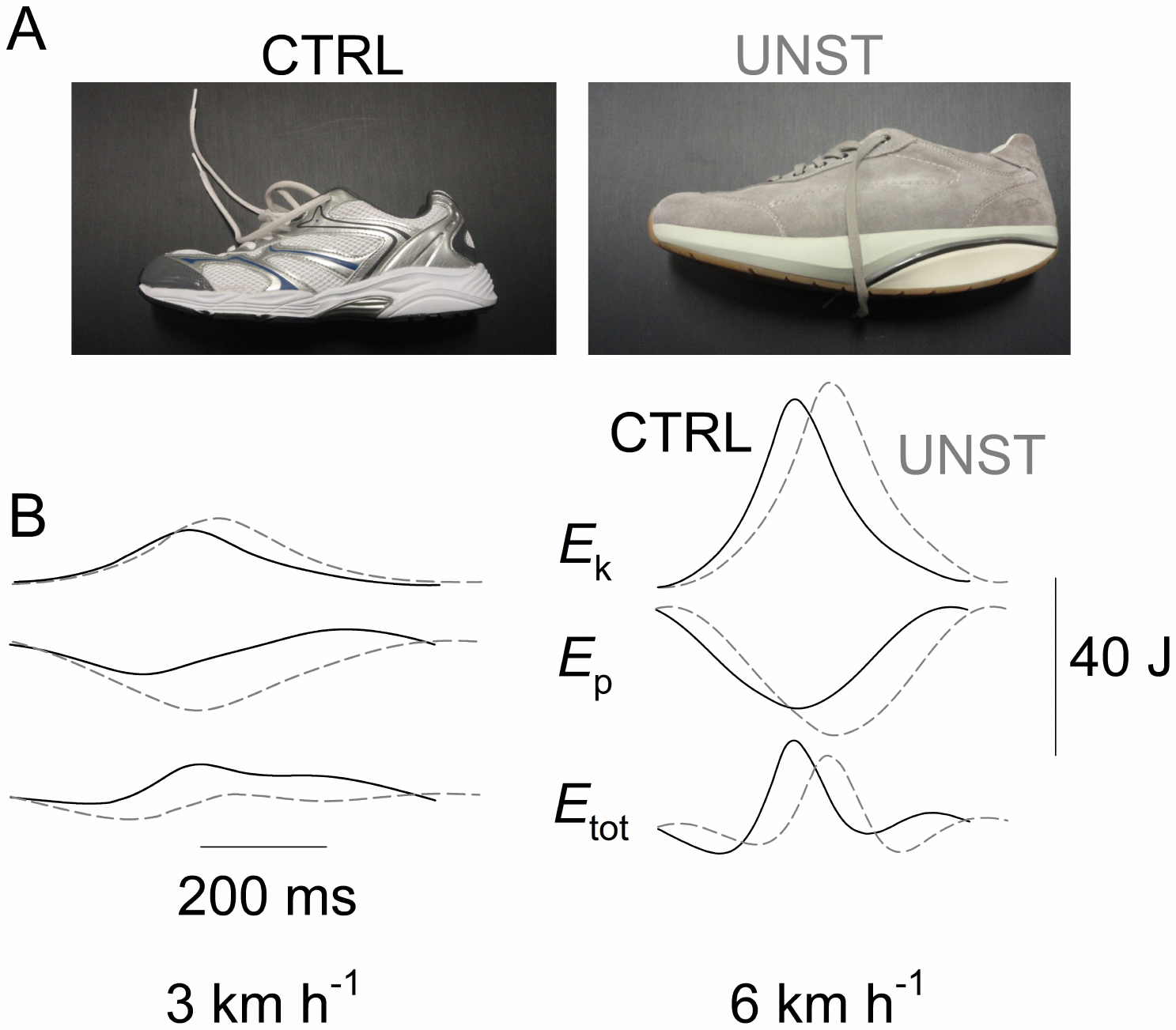}
\caption{Footwear and mechanical energy curves of the CoM during walking. A: Pictures of the CTRL (left) and UNST (right) footwear. B: Typical curves of kinetic (\textit{E}~\hspace{-.1cm}\textsubscript{k}), potential (\textit{E}~\hspace{-.1cm}\textsubscript{p}) and total (\textit{E}~\hspace{-.1cm}\textsubscript{tot}) mechanical energy of the CoM as a function of time at 3 km h\textsuperscript{-1} (left panel) and 6 km h\textsuperscript{-1} (right panel) for CTRL (black solid lines) and UNST (gray dashed lines) footwear.}
\label{Fig. 1}
\end{figure}

Data from force platform were analyzed using the 3D-Force Plate Software (v. 2.7.4, Arsalis, Louvain-la-Neuve, Belgium). Although GRFs were collected in the three directions, the lateral component of GRFs has been neglected since it is irrelevant for the study of forward motion~\cite{bib17}. Starting from the GRFs, the accelerations curves of the CoM are computed from the vertical and forward components of the GRFs and the mass of the participants. The integration of these accelerations gives the vertical and forward velocity curves of the CoM (\textit{V}~\hspace{-.1cm}\textsubscript{v} and \textit{V}~\hspace{-.1cm}\textsubscript{f}), allowing the computation of the CoM kinetic energy (\textit{E}~\hspace{-.1cm}\textsubscript{k}) curve, as well as the vertical displacement (\textit{S}~\hspace{-.1cm}\textsubscript{v}) of the CoM, from which the CoM gravitational potential energy (\textit{E}~\hspace{-.1cm}\textsubscript{p}) curve was computed (see Fig. \ref{Fig. 1}B for a plot of the typical shape of \textit{E}~\hspace{-.1cm}\textsubscript{k} and \textit{E}~\hspace{-.1cm}\textsubscript{p} during a step).

A key point in the data processing is then the identification and the characterization of the different steps. A complete step is defined as the sequence delimited by two consecutive minima of \textit{V}~\hspace{-.1cm}\textsubscript{f}. The time of the step (\textit{T}~\hspace{-.1cm}\textsubscript{step}) is then readily computed, as well as the average progression  speed during a step (\textit{V}~\hspace{-.1cm}\textsubscript{step}). The step length (\textit{L}~\hspace{-.1cm}\textsubscript{step}) is finally given by \textit{L}~\hspace{-.1cm}\textsubscript{step} = \textit{V}~\hspace{-.1cm}\textsubscript{step} \textit{T}~\hspace{-.1cm}\textsubscript{step}.

The external mechanical work (\textit{W}~\hspace{-.1cm}\textsubscript{ext}), i.e. the work performed to accelerate the CoM relative to the ground during a whole step, was computed following the method described in~\cite{bib12,bib18}, summarized here. The total external mechanical energy of the CoM is the sum of \textit{E}~\hspace{-.1cm}\textsubscript{p} and \textit{E}~\hspace{-.1cm}\textsubscript{k}; \textit{E}~\hspace{-.1cm}\textsubscript{tot} = \textit{E}~\hspace{-.1cm}\textsubscript{p} + \textit{E}~\hspace{-.1cm}\textsubscript{k} (see Fig. \ref{Fig. 1}B for a typical plot of \textit{E}~\hspace{-.1cm}\textsubscript{tot}). The sums of the positive increments of \textit{E}~\hspace{-.1cm}\textsubscript{k} (\textit{W}~\hspace{-.1cm}\textsubscript{k} = $\sum\Delta$\textit{E}~\hspace{-.1cm}\textsubscript{k}\textsuperscript{+}) and \textit{E}~\hspace{-.1cm}\textsubscript{p} (\textit{W}~\hspace{-.1cm}\textsubscript{p} = $\sum\Delta$\textit{E}~\hspace{-.1cm}\textsubscript{p}\textsuperscript{+}) represent the positive work necessary to accelerate and lift the CoM respectively and have been calculated for each complete step, together with the total external work (\textit{W}~\hspace{-.1cm}\textsubscript{ext} = $\sum\Delta$\textit{E}~\hspace{-.1cm}\textsubscript{tot}\textsuperscript{+}). Note that \textit{W}~\hspace{-.1cm}\textsubscript{k}, \textit{W}~\hspace{-.1cm}\textsubscript{p} and \textit{W}~\hspace{-.1cm}\textsubscript{ext} are expressed per unit distance (\textit{L}~\hspace{-.1cm}\textsubscript{step}) and weight (participant's mass) in the following plots.

A quantity of interest is also the recovery of mechanical energy of the CoM over the whole step (\textit{R}) defined as~\cite{bib11} $$R = 100\frac{W\textsubscript{p}+W\textsubscript{k}-W\textsubscript{ext}}{W\textsubscript{p}+W\textsubscript{k}}.$$ A 100\% \textit{R} value would mean that the \textit{E}~\hspace{-.1cm}\textsubscript{k} and \textit{E}~\hspace{-.1cm}\textsubscript{p} curves are exactly of opposite phase and of equal shape and amplitude according to a ballistic movement~\cite{bib11}. A value of \textit{R} lower than 100\% means that the maximum of \textit{E}~\hspace{-.1cm}\textsubscript{k} and the minimum of \textit{E}~\hspace{-.1cm}\textsubscript{p} are not simultaneous. The phase shift ($\alpha$) is the temporal difference between the maximum of \textit{E}~\hspace{-.1cm}\textsubscript{k} and minimum of \textit{E}~\hspace{-.1cm}\textsubscript{p}, reported on \textit{T}~\hspace{-.1cm}\textsubscript{step}. It has been measured and further expressed in degrees ($^{\circ}$), knowing that a step is equal to 360~$^{\circ}$.

Finally, the cumulated works \textit{W}~\hspace{-.1cm}\textsubscript{k}, \textit{W}~\hspace{-.1cm}\textsubscript{p} and \textit{W}~\hspace{-.1cm}\textsubscript{ext} have been computed at an instantaneous level. The instantaneous recovery \textit{r}(\textit{t}) and the cumulative recovery \textit{R}~\hspace{-.1cm}\textsubscript{int}(\textit{t}) are defined and computed according to~\cite{bib14}. \textit{R}~\hspace{-.1cm}\textsubscript{int} is the value of \textit{R}~\hspace{-.1cm}\textsubscript{int}(\textit{t}) at the end of a step.

\subsection{Statistical Analysis}
\textit{A priori} required sample size for repeated analysis of variance (RM ANOVA) was computed in the statistical software G*Power (version 3.1.9.2)~\cite{bib19}. A sample size of 22 subjects was required to detect a significant average difference in \textit{W}~\hspace{-.1cm}\textsubscript{ext} with a power of 0.90 and $\alpha$ = 0.05. This calculation was based on a large Cohen's effect size \textit{f} = 0.40 calculated from \textit{W}~\hspace{-.1cm}\textsubscript{ext} results obtained with standard and rocker bottom soles during the single support in a previous study~\cite{bib16}.

All data are presented as means and standard deviations (SD) and were checked for normality (Shapiro-Wilk) and equal variance tests. A two-way (shoe $\times$ speed) RM ANOVA with Holm-Sidak method for pairwise multiple comparisons was performed and used to examine the effect of footwear (2 conditions: CTRL and UNST), speed of progression (4 levels: 3, 4, 5 and 6 km h\textsuperscript{-1}), and interaction effect of footwear and speed of progression on results. The effect size ($\eta$\textsuperscript{2}) was calculated as the sums of the squares for the effect of interest (shoe, speed, speed $\times$ shoe) divided by the total sums of the squares. The significance level $\alpha$ was set at 0.05 for all analyses and \textit{post hoc} statistical power was calculated.

Day to day intra-observer reproducibility of results (24 pairs of observations for each footwear) was examined for investigator GC in a test-retest study. This latter was performed on the mean of 5 walking trials obtained from 6 female subjects included in the study and walking with CTRL or UNST footwear at the same 4 speeds of progression, a day apart and at the same time of day. Reproducibility was estimated using the concordance correlation coefficient (CCC) developed by Lin (1989), that combines measures of both precision ($\rho$) and accuracy (\textit{C}~\hspace{-.1cm}\textsubscript{b}) coefficients to determine how far the test-retest pairs of observations deviate from the line of perfect concordance. A CCC cutoff value of 0.7 was used to indicate test-retest reproducibility.

All statistical procedures were performed with SigmaPlot software version 11.0 (Systat Software, San Jose, CA) and MedCalc software version 13.2.0 (MedCalc Software bvba, Ostend, Belgium).

\section{Results}
Day to day intra-observer reproducibility is presented in Table~\ref{table2}. CCC values for all variables were systematically greater than 0.7 with good to high intra-observer reproducibility for all variables. However, CCC values were specifically smaller for \textit{R}, \textit{R}~\hspace{-.1cm}\textsubscript{int} and $\alpha$ with UNST footwear, and mainly explained by an increase of imprecision ($\rho$ is reduced compared to CTRL). This increased imprecision reflects an increase of variability in the estimates with UNST.

\begin{figure}
\includegraphics[scale=0.6]{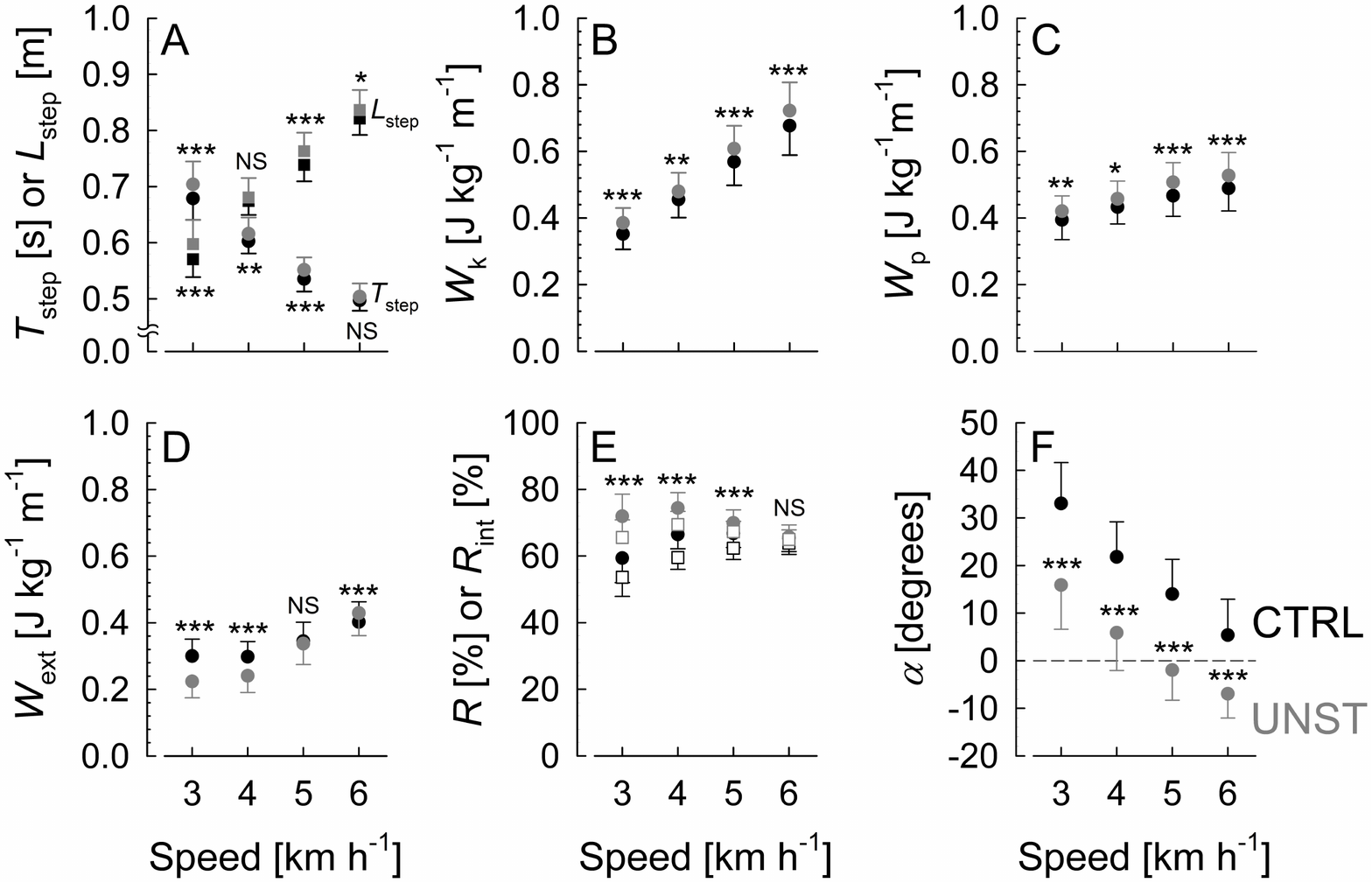}
\caption{Evolution with the speed of the different measured walking variables. Stars are used to denote the level of significance when comparing the data measured with CTRL (black symbols) and UNST footwear (gray symbols): * for p$<$0.05, ** for p$<$0.01, *** for p$<$0.001, and NS (non significant) for p$>$0.05. In panels A and E, squares represent \textit{L}~\hspace{-.1cm}\textsubscript{step} and \textit{R} respectively.}
\label{Fig. 2}
\end{figure}

The values of \textit{L}~\hspace{-.1cm}\textsubscript{step} and \textit{T}~\hspace{-.1cm}\textsubscript{step} are significantly influenced by speed of progression and footwear (Fig. \ref{Fig. 2}A, Table~\ref{table3}) with a significantly higher step length for UNST compared to CTRL footwear at 3, 5, and 6 km h\textsuperscript{-1} (Table~\ref{table4}) and a significantly longer time of step for UNST compared to CTRL footwear at 3, 4, and 5 km h\textsuperscript{-1} (Table~\ref{table4}).

Both \textit{W}~\hspace{-.1cm}\textsubscript{k} and \textit{W}~\hspace{-.1cm}\textsubscript{p} are significantly influenced by speed of progression and footwear (Fig. \ref{Fig. 2}B and Fig. \ref{Fig. 2}C, Table~\ref{table3}). \textit{W}~\hspace{-.1cm}\textsubscript{k} and \textit{W}~\hspace{-.1cm}\textsubscript{p} values are significantly increased for UNST compared to CTRL footwear, whatever the speed of progression (Table~\ref{table4}). \textit{W}~\hspace{-.1cm}\textsubscript{ext} values are significantly influenced by speed of progression and footwear (Fig. \ref{Fig. 2}D, Table~\ref{table3}). With UNST footwear, \textit{W}~\hspace{-.1cm}\textsubscript{ext} values were nearly constant at 3  and 4 km h\textsuperscript{-1}. To higher speeds of progression, it increases abruptly to reach a maximal value at 6 km h\textsuperscript{-1}. When comparing \textit{W}~\hspace{-.1cm}\textsubscript{ext} values with CTRL and UNST footwear, a more complex behavior than in the cases of \textit{W}~\hspace{-.1cm}\textsubscript{k} and \textit{W}~\hspace{-.1cm}\textsubscript{p} is observed. First, at low speeds of progression, \textit{W}~\hspace{-.1cm}\textsubscript{ext} values are significantly lower for UNST compared to CTRL footwear. Second, at an intermediate speed of 5 km h\textsuperscript{-1}, no significant difference of \textit{W}~\hspace{-.1cm}\textsubscript{ext} is observed. Third, at a higher speed of 6 km h\textsuperscript{-1}, \textit{W}~\hspace{-.1cm}\textsubscript{ext} becomes significantly higher for UNST compared to CTRL footwear.

\begin{figure}
\includegraphics[scale=0.7]{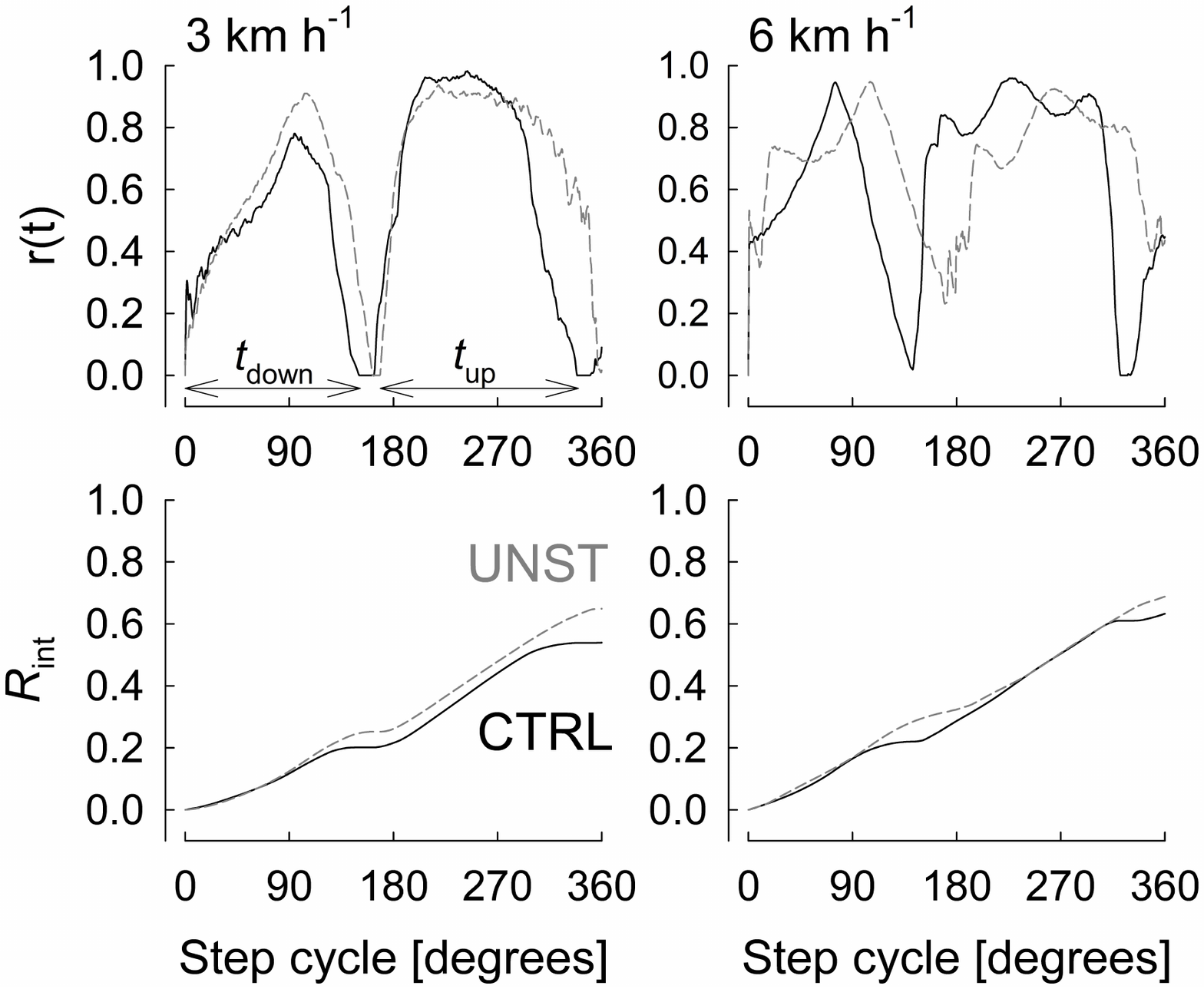}
\caption{ Typical representations of instantaneous recovery (\textit{r}(\textit{t})) and cumulative recovery (\textit{R}~\hspace{-.1cm}\textsubscript{int}) for UNST (gray dashed lines) and CTRL footwear (black solid lines) as a function of the step cycle, at 3 km  h\textsuperscript{-1} (left panel) and 6 km  h\textsuperscript{-1} (right panel). The time is expressed in degrees, where 0~$^{\circ}$ and 360~$^{\circ}$ correspond to the beginning and the end of a step respectively.}
\label{Fig. 3}
\end{figure}

Results of \textit{R} and \textit{R}~\hspace{-.1cm}\textsubscript{int} as function of speed of progression are shown in Fig. \ref{Fig. 2}E, while instantaneous values for \textit{r}(\textit{t}) and \textit{R}~\hspace{-.1cm}\textsubscript{int}(\textit{t}) are shown in Fig. \ref{Fig. 3}. Although the values are slightly different, a similar trend is observed for \textit{R} and \textit{R}~\hspace{-.1cm}\textsubscript{int}. Both \textit{R} and \textit{R}~\hspace{-.1cm}\textsubscript{int} values are significantly influenced by speed of progression and footwear (Table~\ref{table3}). \textit{R} and \textit{R}~\hspace{-.1cm}\textsubscript{int} values are significantly higher for UNST compared to CTRL footwear at 3, 4, and 5 km h\textsuperscript{-1} (Table~\ref{table4}). \textit{R}~\hspace{-.1cm}\textsubscript{int} is increasing during the step cycle, by definition (Fig. \ref{Fig. 3}). Nevertheless, some plateaus can be observed, corresponding to periods where \textit{W}~\hspace{-.1cm}\textsubscript{ext} is maximal. The increase of \textit{R}~\hspace{-.1cm}\textsubscript{int} shows two plateaus during the step, the first at around 180~$^{\circ}$ and the second at around 360~$^{\circ}$. The first plateau takes place during the double contact and the second one takes place during the swing at the end of the step. For both types of footwear, the plateaus are smaller as speed increases. However, UNST footwear leads to shorter plateaus than CTRL footwear. This is coherent with the fact that \textit{r}(\textit{t}) is null on shorter time intervals with UNST footwear than with CTRL footwear ; at 6 km h\textsuperscript{-1} it may even happen than \textit{r}(\textit{t}) is always greater than zero with UNST footwear (Fig. \ref{Fig. 3}).

The phase shift is plotted in Fig. \ref{Fig. 2}F. It is a decreasing function of the speed for both types of footwear and can become negative at high speed with UNST footwear. When comparing both types of footwear, it appears that $\alpha$ is always significantly lower for UNST footwear than for CTRL footwear (Table~\ref{table4}). At 3 km  h\textsuperscript{-1} with UNST footwear, \textit{r}(\textit{t}) is greater during the descent of the CoM (\textit{t}~\hspace{-.1cm}\textsubscript{down}) and at the end of the ascent (\textit{t}~\hspace{-.1cm}\textsubscript{up}) compared to CTRL, resulting in a significant net increase in \textit{R}~\hspace{-.1cm}\textsubscript{int}. At 6 km  h\textsuperscript{-1}, this increase is much less marked and is non significant (see Table~\ref{table4}).


\section{Discussion}

The main findings of the present study were that commercially available unstable footwear equipment could introduce noise during walking since it increases variability and modifies the pendulum energy conversion during (\textit{R}) and within (\textit{R}~\hspace{-.1cm}\textsubscript{int}) the step cycle and the resultant \textit{W}~\hspace{-.1cm}\textsubscript{ext} performed in healthy young women after the first 30 min of exposure. Since most of unstable shoes on the market have a rounded sole design, like MBT shoes, we believe that our findings also apply to other models with rounded soles like Skechers Shape-Ups.

Since skilled performance is typically associated with a low level of variability~\cite{bib21} and that reproducibility of spatiotemporal variables of the step were similar for both CTRL and UNST footwear
, we do not believe that our results are influenced by a short-term learning process and thus accurately reflect the immediate adaptation of subjects with UNST footwear. However, we cannot exclude the existence of a learning process over a longer term that could be identified only through a longitudinal study design for months or even years. A previous study explored the effects of a 10-week training period with UNST footwear on several biomechanical variables~\cite{bib2} but pendulum energy conversion and \textit{W}~\hspace{-.1cm}\textsubscript{ext} were not estimated. 

 The major differences between UNST and CTRL footwear are: (1) a decreased radius of curvature of sole, (2) an increased thickness of sole and total weight, and (3) a decreased ratio of flat to total area of the sole. Let us discuss the influence of these three factors.

First, the effects of arc-foot length and radius on CoM velocity fluctuations were previously estimated~\cite{bib22} and it has been shown that length has much greater effect on the mechanical work of the step-to-step transition than radius that has a comparatively subtler effect on walking. A very high radius of curvature, like a flat foot bottom as our CTRL footwear, may be quite unfavorable since the center of pressure advance very rapidly, requiring compensatory forces by the muscles to be produced~\cite{bib22}. This is in accordance with our \textit{W}~\hspace{-.1cm}\textsubscript{ext} results showing that more work is done by the musculoskeletal system when walking at low speed with CTRL footwear. In this study, since the same size was used for CTRL and UNST footwear, the total shoe length are nearly similar (difference of 7 mm) and arc-foot length could not influence our results.

Second, since the thickness of the sole was greater for UNST footwear than CTRL, the lower limb length was artificially increased. 
However, we do not believe that a small thickness difference of 1.5 cm at midfoot could significantly influence the temporal stride kinematics (the average size of our population is 167 cm). Rather, the difference in weight between UNST and CTRL footwear is the main source of increased step length. Even if foot loads principally increase the metabolic rate required for swinging the leg, a 4-kg foot load increased stride length by 6\%~\cite{bib23}. Our findings indicate a step length increase of 1.9 to 4.7\% for a footwear load difference of 400 g. In addition, our results are also in good agreement with those of~\cite{bib24} who observed an increase (decrease) of \textit{L}~\hspace{-.1cm}\textsubscript{step} (\textit{T}~\hspace{-.1cm}\textsubscript{step}) except for \textit{T}~\hspace{-.1cm}\textsubscript{step} where they find a significant difference for 5 km h\textsuperscript{-1} and not for 6 km h\textsuperscript{-1}, while we reach the opposite conclusion. These differences could be attributed to the treadmill walking used in this study instead of level walking and that we accepted walking trials at target speed at $\pm$ 0.3 km h\textsuperscript{-1}. At a given speed, since a longer \textit{L}~\hspace{-.1cm}\textsubscript{step} was adopted when wearing UNST, an increased vertical displacement of the CoM and \textit{W}~\hspace{-.1cm}\textsubscript{p} was theoretically expected. Our results show that \textit{W}~\hspace{-.1cm}\textsubscript{p} was significantly increased at all speeds when wearing UNST. A second consequence of the longer \textit{L}~\hspace{-.1cm}\textsubscript{step} is that greater variations in \textit{V}~\hspace{-.1cm}\textsubscript{f} curve of the CoM and an increased \textit{W}~\hspace{-.1cm}\textsubscript{k} were expected, as a result of an increased breaking of the leading limb with the ground in the forward direction. Again, our results show that \textit{W}~\hspace{-.1cm}\textsubscript{k} was significantly increased at all speeds when wearing UNST. 

Third, a shorter flat region of sole in UNST footwear increased the efficiency of pendular energy conversion. This phenomenon was expected since a more curved geometry of sole must allow the body CoM to ``roll" forward, taking advantage of the passive dynamics of a rocker-based inverted pendulum~\cite{bib25}. However, this phenomenon was only observed at slow walking speeds. MBT shoes thus facilitate walking at low speed from a mechanical point of view, while it shows no significant influence at higher speed. 

To the best of our knowledge, only one study examined the effects of wearing commercially available rocker-bottom footwear (MBT) on \textit{W}~\hspace{-.1cm}\textsubscript{ext} performed while walking~\cite{bib16}. However, this study does not explore the pendulum energy conversion during or within the step cycle nor other walking speed than the one spontaneously adopted by their subjects.
 Our results are in good agreement with those of this study since no significant difference in \textit{W}~\hspace{-.1cm}\textsubscript{ext} was observed between UNST and CTRL footwear at a speed of 5 km h\textsuperscript{-1}.

The main strength of our study is to have investigated other walking speeds than the spontaneous one. This allowed us to demonstrate that UNST footwear has a greater influence on \textit{W}~\hspace{-.1cm}\textsubscript{ext} than previously thought, with not only an earlier redistribution in the step cycle~\cite{bib16}. On the contrary, our findings indicate that walking speed mainly influences \textit{W}~\hspace{-.1cm}\textsubscript{ext}, and that this latter could be unchanged, increased or decreased as compared to CTRL footwear. At the preferred human walking speed of 5 km h\textsuperscript{-1}, \textit{W}~\hspace{-.1cm}\textsubscript{ext} with UNST rocker-bottom footwear was unchanged compared to CTRL. At walking speeds lower than 5 km h\textsuperscript{-1}, UNST footwear requires lesser \textit{W}~\hspace{-.1cm}\textsubscript{ext} to be performed; while at 6 km h\textsuperscript{-1} more \textit{W}~\hspace{-.1cm}\textsubscript{ext} was performed. This discovery allows us to qualify the statement that UNST could serve as a motor constraint process~\cite{bib2}. We conclude that UNST rocker-bottom footwear could be used as a motor constraint at brisk speed but as a motor facilitator at slow speed, at least in terms of \textit{W}~\hspace{-.1cm}\textsubscript{ext}. In the same vein, our findings also show that the efficiency of pendulum energy conversion is superior with UNST footwear at slow speed, indicating that the dynamics of walking is then closer to a perfect pendular motion. 
From a mechanical viewpoint, such higher \textit{R}-values originate in lower potential-kinetic energy curves dephasing time for UNST footwear. Cavagna et al.~\cite{bib14} developed an analytical model linking CoM energy curves phase shift to \textit{R} and \textit{R}~\hspace{-.1cm}\textsubscript{int}. In this mode, opposite values for the phase shift lead to the same value of \textit{R}~\hspace{-.1cm}\textsubscript{int}. Our experimental findings are in accordance with that since \textit{R} and \textit{R}~\hspace{-.1cm}\textsubscript{int} values do not differ significantly at 6 km h\textsuperscript{-1}, a brisk speed at which nearly opposite $\alpha$ values of the phase shift are found for both types of shoes. 

Our findings support the idea that unstable rocker-bottom footwear could serve as a speed-dependent noise-based training gear to exercise pendulum energy conversion during walking in healthy young women. At preferred walking speed, this equipment does not modify \textit{W}~\hspace{-.1cm}\textsubscript{ext}. At slow speed, it acts as a stochastic resonance or locomotor facilitator by improving the efficiency of pendulum energy conversion and reducing \textit{W}~\hspace{-.1cm}\textsubscript{ext}, whereas at brisk speed it acts as a locomotor constraint by decreasing efficiency of conversion and increasing \textit{W}~\hspace{-.1cm}\textsubscript{ext}. However, further studies on pendulum energy conversion during and within the step cycle with unstable rocker-bottom footwear at speeds at the upper end of the natural range of walking, typically 7 to 9 km  h\textsuperscript{-1}, are needed to complete our results. 

 

\newpage

\begin{table}[!h]
\begin{adjustwidth}{-0.5in}{0in} 
\caption{Characteristics of footwear. The radius of curvature of the sole is the radius of the circle that fits at best the sole's surface. Sole's flat area is the sole's contact area with the ground in neutral position.}
\begin{tabular}{llll}
\hline
                             &            & CTRL (Osaga) &      UNST (MBT)\\ \hline
                                      
Radius of curvature of sole (cm) & Forefoot        & 26.9        & 33.0            \\
      	& Midfoot & 41.0 & 25.6  \\
      	& Rearfoot & 24.3 & 18.8 \\
                              &    Mean$\pm$SD            & 30.7$\pm$7.3 &        25.8$\pm$5.8                   \\
Outsole height relative to ground (cm)  & Forefoot & 4       & 3         \\
 & Midfoot & 2.5  & 4 \\
 & Rearfoot & 4.5  & 6 \\
                                 &  Mean$\pm$SD             & 3.7$\pm$0.8  &     4.3$\pm$1.2          \\
Size (EU/US/UK)        &                  &          40/7/6   &        40/7/6          \\
Weight (g)                             &              & 240      &   440            \\
Ratio of flat to total area of sole (\%) &              & 47       &   18           \\
Total length (cm)                         &              & 28.2     &      27.5    \\ \hline        
\end{tabular}
\label{table1}
\end{adjustwidth}
\end{table}

\newpage

\begin{table}[!h]
\caption{
Reproducibility results for flat walking with control (CTRL) and rocker bottom unstable (UNST) footwear. CCC: concordance correlation coefficient; CI: confidence interval. }
\begin{tabular}{llllllll}
\hline
                      & CCC   & 95\% CI     & Precision ($\rho$) & Accuracy (\textit{C}~\hspace{-.1cm}\textsubscript{b}) \\ \hline
CTRL                  &       &             &                &               \\ \hline
\textit{V}~\hspace{-.1cm}\textsubscript{step}     & 0.991 & 0.979--0.996 & 0.991          & 0.999         \\
\textit{T}~\hspace{-.1cm}\textsubscript{step}        & 0.951 & 0.891--0.978 & 0.952          & 0.999         \\
\textit{L}~\hspace{-.1cm}\textsubscript{step}        & 0.933 & 0.853--0.971 & 0.937          & 0.996         \\
\textit{W}~\hspace{-.1cm}\textsubscript{ext}      & 0.841 & 0.687--0.923 & 0.863          & 0.975         \\
\textit{W}~\hspace{-.1cm}\textsubscript{k}      & 0.966 & 0.927--0.985 & 0.970           & 0.996         \\
\textit{W}~\hspace{-.1cm}\textsubscript{p}      & 0.770  & 0.552--0.889 & 0.809          & 0.952         \\
\textit{R}            & 0.853 & 0.705--0.930 & 0.871          & 0.979         \\
\textit{R}~\hspace{-.1cm}\textsubscript{int}        & 0.930  & 0.850--0.969 & 0.935          & 0.995         \\
$\alpha$      & 0.879 & 0.749--0.944 & 0.888          & 0.990          \\
                      &       &             &                &               \\ \hline
UNST                  &       &             &                &               \\ \hline
\textit{V}~\hspace{-.1cm}\textsubscript{step}    & 0.996 & 0.991--0.998 & 0.996          & 0.999         \\
\textit{T}~\hspace{-.1cm}\textsubscript{step}        & 0.975 & 0.945--0.989 & 0.979          & 0.996         \\
\textit{L}~\hspace{-.1cm}\textsubscript{step}        & 0.955 & 0.901--0.980 & 0.958          & 0.998         \\
\textit{W}~\hspace{-.1cm}\textsubscript{ext}  & 0.939 & 0.865--0.973 & 0.940           & 0.999         \\
\textit{W}~\hspace{-.1cm}\textsubscript{k}   & 0.961 & 0.913--0.982 & 0.963          & 0.998         \\
\textit{W}~\hspace{-.1cm}\textsubscript{p}    & 0.715 & 0.483--0.853 & 0.778          & 0.918         \\
\textit{R}             & 0.756 & 0.542--0.878 & 0.787          & 0.960          \\
\textit{R}~\hspace{-.1cm}\textsubscript{int}        & 0.703 & 0.468--0.845 & 0.754          & 0.932         \\
$\alpha$       & 0.737 & 0.486--0.876 & 0.743          & 0.992     \\ \hline   
\end{tabular}

\label{table2}
\end{table}

\newpage

\begin{table}[!h]
\caption{ Two-way RM ANOVA results.}
\begin{tabular}{llllll}
\hline
                      & \textit{df} & F      & P              & Effect size ($\eta$\textsuperscript{2}) & Power \\ \hline
\textit{L}~\hspace{-.1cm}\textsubscript{step} {[}m{]}         &    &        &                &                  &       \\ \hline
Speed                 & 3  & 556.05 & \textless0.001 & 0.88             & 1     \\
Footwear              & 1  & 29.84  & \textless0.001 & 0.02             & 1     \\
Speed $\times$ Footwear      & 3  & 2.62   & 0.058          & 0.002            & 0.4   \\ \hline
\textit{T}~\hspace{-.1cm}\textsubscript{step} {[}s{]}         &    &        &                &                  &       \\ \hline
Speed                 & 3  & 444.51 & \textless0.001 & 0.86             & 1     \\
Footwear              & 1  & 31.48  & \textless0.001 & 0.01             & 1     \\
Speed $\times$ Footwear      & 3  & 2.95   & 0.038          & 0.002            & 0.47  \\ \hline
\textit{W}~\hspace{-.1cm}\textsubscript{k} {[}J kg\textsuperscript{-1} m\textsuperscript{-1}{]}   &    &        &                &                  &       \\ \hline
Speed                 & 3  & 457.65 & \textless0.001 & 0.77             & 1     \\
Footwear              & 1  & 43     & \textless0.001 & 0.02             & 1     \\
Speed $\times$ Footwear      & 3  & 2.11   & 0.107          & 0.001            & 0.28  \\ \hline
\textit{W}~\hspace{-.1cm}\textsubscript{p} {[}J kg\textsuperscript{-1} m\textsuperscript{-1}{]}   &    &        &                &                  &       \\ \hline
Speed                 & 3  & 39.38  & \textless0.001 & 0.29             & 1     \\
Footwear              & 1  & 19.33  & \textless0.001 & 0.05             & 0.99  \\
Speed $\times$ Footwear      & 3  & 1.35   & 0.264          & 0.003            & 0.11  \\ \hline
\textit{W}~\hspace{-.1cm}\textsubscript{ext} {[}J kg\textsuperscript{-1} m\textsuperscript{-1}{]} &    &        &                &                  &       \\ \hline
Speed                 & 3  & 100.54 & \textless0.001 & 0.52             & 1     \\
Footwear              & 1  & 35.35  & \textless0.001 & 0.03             & 1     \\
Speed $\times$ Footwear      & 3  & 54.2   & \textless0.001 & 0.02             & 1     \\ \hline
\textit{R} {[}\%{]}            &    &        &                &                  &       \\ \hline
Speed                 & 3  & 177.55 & \textless0.001 & 0.1              & 1     \\
Footwear              & 1  & 13.24  & \textless0.001 & 0.21             & 1     \\
Speed $\times$ Footwear      & 3  & 62.48  & \textless0.001 & 0.13             & 1     \\ \hline
\textit{R}~\hspace{-.1cm}\textsubscript{int} {[}\%{]}         &    &        &                &                  &       \\ \hline
Speed                 & 3  & 26.62  & \textless0.001 & 0.13             & 1     \\
Footwear              & 1  & 204.09 & \textless0.001 & 0.33             & 1     \\
Speed $\times$ Footwear      & 3  & 65.18  & \textless0.001 & 0.12             & 1     \\ \hline
$\alpha$ {[}$^{\circ}${]}      &    &        &                &                  &       \\ \hline
Speed                 & 3  & 120.18 & \textless0.001 & 0.44             & 1     \\
Footwear              & 1  & 271.52 & \textless0.001 & 0.29             & 1     \\
Speed $\times$ Footwear      & 3  & 4.12   & 0.01           & 0.004            & 0.7  \\ \hline

\end{tabular}
\label{table3}
\end{table}

\newpage

\begin{table}[!h]
\caption{
Post-hoc Holm-Sidak results for footwear factor. CTRL: control footwear, UNST: unstable footwear.}
\begin{tabular}{llllll}
\hline

                      & CTRL       & UNST        & Difference & t     & P              \\
                      & (mean$\pm$SD)  & (mean$\pm$SD)   & of means   &       &                \\ \hline
\textit{L}~\hspace{-.1cm}\textsubscript{step} {[}m{]}         &            &             &            &       &                \\ \hline
3 km h\textsuperscript{-1}              & 0.571$\pm$0.03 & 0.598$\pm$0.04  & 0.027      & 4.46  & \textless0.001 \\
4 km h\textsuperscript{-1}              & 0.675$\pm$0.02 & 0.681$\pm$0.03  & 0.006      & 1.02  & 0.312          \\
5 km h\textsuperscript{-1}              & 0.739$\pm$0.03 & 0.763$\pm$0.03  & 0.024      & 4.01  & \textless0.001 \\
6 km h\textsuperscript{-1}              & 0.821$\pm$0.03 & 0.837$\pm$0.04  & 0.015      & 2.54  & 0.013          \\ \hline
\textit{T}~\hspace{-.1cm}\textsubscript{step} {[}s{]}         &            &             &            &       &                \\ \hline
3 km h\textsuperscript{-1}              & 0.679$\pm$0.04 & 0.704$\pm$0.04  & 0.025      & 5.28  & \textless0.001 \\
4 km h\textsuperscript{-1}              & 0.603$\pm$0.02 & 0.616$\pm$0.03  & 0.013      & 2.71  & 0.008          \\
5 km h\textsuperscript{-1}              & 0.536$\pm$0.02 & 0.552$\pm$0.016 & 0.016      & 3.37  & 0.001          \\
6 km h\textsuperscript{-1}              & 0.498$\pm$0.02 & 0.504$\pm$0.02  & 0.006      & 1.34  & 0.184          \\ \hline
\textit{W}~\hspace{-.1cm}\textsubscript{k} {[}J kg\textsuperscript{-1} m\textsuperscript{-1}{]}   &            &             &            &       &                \\ \hline
3 km h\textsuperscript{-1}              & 0.352$\pm$0.05 & 0.385$\pm$0.05  & 0.033      & 4.38  & \textless0.001 \\
4 km h\textsuperscript{-1}              & 0.456$\pm$0.05 & 0.479$\pm$0.06  & 0.023      & 3.08  & 0.003          \\
5 km h\textsuperscript{-1}              & 0.569$\pm$0.07 & 0.608$\pm$0.07  & 0.039      & 5.12  & \textless0.001 \\
6 km h\textsuperscript{-1}              & 0.677$\pm$0.09 & 0.722$\pm$0.08  & 0.045      & 5.88  & \textless0.001 \\ \hline
\textit{W}~\hspace{-.1cm}\textsubscript{p} {[}J kg\textsuperscript{-1} m\textsuperscript{-1}{]}   &            &             &            &       &                \\ \hline
3 km h\textsuperscript{-1}              & 0.394$\pm$0.06 & 0.421$\pm$0.05  & 0.026      & 2.68  & 0.009          \\
4 km h\textsuperscript{-1}              & 0.433$\pm$0.05 & 0.457$\pm$0.05  & 0.024      & 2.49  & 0.015          \\
5 km h\textsuperscript{-1}              & 0.467$\pm$0.06 & 0.508$\pm$0.06  & 0.041      & 4.2   & \textless0.001 \\
6 km h\textsuperscript{-1}              & 0.489$\pm$0.07 & 0.527$\pm$0.07  & 0.038      & 3.91  & \textless0.001 \\ \hline
\textit{W}~\hspace{-.1cm}\textsubscript{ext} {[}J kg\textsuperscript{-1} m\textsuperscript{-1}{]} &            &             &            &       &                \\ \hline
3 km h\textsuperscript{-1}              & 0.300$\pm$0.05 & 0.224$\pm$0.05  & 0.076      & 10.44 & \textless0.001 \\
4 km h\textsuperscript{-1}              & 0.298$\pm$0.04 & 0.241$\pm$0.05  & 0.057      & 7.83  & \textless0.001 \\
5 km h\textsuperscript{-1}              & 0.344$\pm$0.06 & 0.337$\pm$0.06  & 0.007      & 0.96  & 0.34           \\
6 km h\textsuperscript{-1}              & 0.403$\pm$0.06 & 0.429$\pm$0.07  & 0.026      & 3.61  & \textless0.001 \\ \hline
\textit{R} {[}\%{]}            &            &             &            &       &                \\ \hline
3 km h\textsuperscript{-1}              & 59.37$\pm$7.38 & 71.91$\pm$6.63  & 12.688     & 16.89 & \textless0.001 \\
4 km h\textsuperscript{-1}              & 66.48$\pm$4.30 & 74.41$\pm$4.62  & 7.982      & 10.62 & \textless0.001 \\
5 km h\textsuperscript{-1}              & 66.77$\pm$4.22 & 69.93$\pm$3.94  & 3.124      & 4.16  & \textless0.001 \\
6 km h\textsuperscript{-1}              & 65.44$\pm$4.12 & 65.72$\pm$3.60  & 0.223      & 0.29  & 0.768          \\ \hline
\textit{R}~\hspace{-.1cm}\textsubscript{int} {[}\%{]}         &            &             &            &       &                \\ \hline
3 km h\textsuperscript{-1}              & 53.63$\pm$5.77 & 65.54$\pm$5.30  & 11.911     & 16.66 & \textless0.001 \\
4 km h\textsuperscript{-1}              & 59.53$\pm$3.54 & 69.40$\pm$3.98  & 9.872      & 13.81 & \textless0.001 \\
5 km h\textsuperscript{-1}              & 62.35$\pm$3.41 & 67.31$\pm$3.08  & 4.960       & 6.95  & \textless0.001 \\
6 km h\textsuperscript{-1}              & 63.84$\pm$3.38 & 64.95$\pm$2.89  & 1.106      & 1.55  & 0.126          \\ \hline
$\alpha$ {[}$^{\circ}${]}      &            &             &            &       &                \\ \hline
3 km h\textsuperscript{-1}              & 33.03$\pm$8.60 & 15.89$\pm$9.29  & 17.144     & 13.3  & \textless0.001 \\
4 km h\textsuperscript{-1}              & 21.79$\pm$7.38 & 5.86$\pm$7.91   & 15.925     & 12.36 & \textless0.001 \\
5 km h\textsuperscript{-1}              & 14.00$\pm$7.31 & -1.97$\pm$6.33  & 15.977     & 12.4  & \textless0.001 \\
6 km h\textsuperscript{-1}              & 5.38$\pm$7.52  & -6.95$\pm$5.08  & 12.329     & 9.57  & \textless0.001 \\ \hline

\end{tabular}

\label{table4}
\end{table}

\newpage

\end{document}